\pdfoutput=1 
\documentclass[a4paper,11pt]{article}
\usepackage{jinstpub} 

\begin{document}

\title{Characterization of Silicon Photomultiplier Photon Detection Efficiency at Liquid Nitrogen Temperature}

\newcommand{\uw}{Center for Experimental Nuclear Physics and Astrophysics, and Department of Physics, University of Washington, Seattle, WA 98195, USA}
\newcommand{\iu}{Center for Exploration of Energy and Matter, and Department of Physics, Indiana University, Bloomington, IN 47405, USA}

\newcommand{\orcid}[1]{\href{https://orcid.org/#1}{\includegraphics[width=8pt]{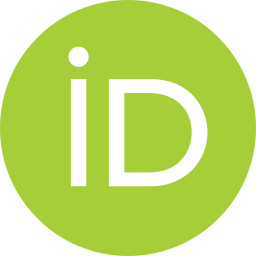}}}

\author[a]{S.~Borden\, \orcid{0009-0003-2539-4333}}
\author[a]{J.A.~Detwiler\, \orcid{0000-0002-9050-4610}}
\author[b]{W.~Pettus\,\orcid{0000-0003-4947-7400}}
\author[a]{N.W.~Ruof\, \orcid{0000-0001-9665-6722}}

\affiliation[a]{\uw}
\affiliation[b]{\iu}

\emailAdd{sjborden@uw.edu}

\abstract{The detection of individual photons at cryogenic temperatures is of interest to many experiments searching for physics beyond the Standard Model. Silicon photomultipliers (SiPMs) are often deployed in liquid argon or liquid xenon to detect scintillation light either directly or after it has been wavelength-shifted. Maximizing the photon detection efficiency (PDE) of the SiPMs used in these experiments optimizes the sensitivity to new physics; however, the PDEs of commercial SiPMs, although well known at room temperature, are not well characterized at the cryogenic temperatures at which many experiments operate them. Here we present results from an experimental setup that measures the photon detection efficiencies of silicon photomultipliers at liquid nitrogen temperature, 77~K. Results from a KETEK PM3325-WB-D0 and a Hamamatsu S13360-3050CS silicon photomultiplier --- of R\&D interest to the LEGEND experiment --- exhibit a decrease in photon detection efficiency greater than 20\% at liquid nitrogen temperature relative to room temperature for $562$~nm light.}


\maketitle
\flushbottom

\section{Introduction}
Silicon photomultipliers (SiPMs) are solid-state detectors with single-photon resolution and can be used to determine the arrival time of individual photons with sub-nanosecond precision due to their good single photon time resolutions and good photon detection efficiencies \cite{Piemonte_Gola_2019, Betancourt_Datwyler_Owen_Puig_Serra_2020, Gundacker_MartinezTurtos_Kratochwil_Pots_Paganoni_Lecoq_Auffray_2020, Nemallapudi_Gundacker_Lecoq_Auffray_2016, Gundacker_Heering_2020}.  This makes SiPMs ideal candidates for low-light detection applications \cite{acerbiUnderstandingSimulatingSiPMs2019, Renker_Lorenz_2009, Klanner_2019, Piemonte_Gola_2019, Buzhan_Dolgoshein_Filatov_Ilyin_Kantzerov_Kaplin_Karakash_Kayumov_Klemin_Popova_etal_2003, Gundacker_Heering_2020}.  A SiPM contains an array of Single Photon Avalanche Diodes (SPADs), each containing a quenching resistor, and is operated a few volts above the breakdown voltage. An incident photon can be absorbed within a SPAD and trigger a self-sustaining avalanche of charge carriers with some probability. Several rare-event search experiments have operated SiPMs at cryogenic temperatures in liquid argon \cite{Agostini_Bakalyarov_Balata_Barabanov_Baudis_Bauer_Bellotti_Belogurov_Belyaev_Benato_etal._2018, Schwarz_Krause_Leonhardt_Papp_Schonert_Wiesinger_Fomina_Gusev_Rumyantseva_Shevchik_etal._2021, Protodune, Agnes_Albergo_Albuquerque_Arba_Ave_Boiano_Bonivento_Bottino_Bussino_Cadeddu_etal._2021, Aalseth_Abdelhakim_Agnes_Ajaj_Albuquerque_Alexander_Alici_Alton_Amaudruz_Ameli_etal._2021} and liquid xenon \cite{Chiappini_2023}, or are planning on operating with SiPMs optimized to detect the scintillation light from liquid argon \cite{Falcone_Andreani_Bertolucci_Brizzolari_Buckanam_Capasso_Cattadori_Carniti_Citterio_Francis_etal._2021, Aalseth_Acerbi_Agnes_Albuquerque_Alexander_Alici_Alton_Antonioli_Arcelli_Ardito_etal._2018, Kochanek_2020} and liquid xenon \cite{Jamil_Ziegler_Hufschmidt_Li_Lupin-Jimenez_Michel_Ostrovskiy_Retiere_Schneider_Wagenpfeil_etal._2018, darwin, pelato}.

The work here is motivated by LEGEND --- an experiment searching for neutrinoless double-beta ($0\nu\beta\beta$) decay in $^{76}$Ge --- because it uses SiPMs for the readout of its liquid argon (LAr) scintillation detector \cite{abgrallLargeEnrichedGermanium2017, legendcollaborationLEGEND1000PreconceptualDesign2021}. The detection of argon scintillation in coincidence with a germanium detector energy deposition is used to tag and remove events that do not resemble $0\nu\beta\beta$ decay, which should be contained entirely within the germanium. Silicon photomultiplers were chosen as the main light readout technology for LEGEND-200 over photomultiplier tubes due to their low radioactive background, low operating voltage, and comparably high quantum efficiency \cite{gerda_sipm_rnd, Schwarz_Krause_Leonhardt_Papp_Schonert_Wiesinger_Fomina_Gusev_Rumyantseva_Shevchik_etal._2021, Krause_thesis}. The liquid argon scintillation, peaked at 128 nm in the vacuum ultraviolet (VUV), is wavelength-shifted and guided to several arrays of SiPMs by a shroud of tetraphenyl butadiene (TPB) coated, wavelength-shifting (WLS) optical fibers \cite{Schwarz_Krause_Leonhardt_Papp_Schonert_Wiesinger_Fomina_Gusev_Rumyantseva_Shevchik_etal._2021}. There are two wavelength conversions involved in this collection process: the TPB first shifts the VUV scintillation light to approximately 430~nm \cite{Kuzniak_Szelc_2020}, and then the WLS optical fibers shift the light to green wavelengths, approximately 460~nm to 580~nm. These green light wavelengths (460~nm to 580~nm) were chosen because they are well matched to the spectral response of the KETEK PM33100T SiPMs used for light readout in LEGEND-200 \cite{Wiesinger}, as well as other models of SiPMs such as the RGB-HD from Fondazione Bruno Kessler, the S13360 from Hamamatsu, or any other commercially available visible light sensitive SiPMs. This method of collecting the VUV scintillation offers a large area of exposure to scintillation as opposed to using VUV-sensitive SiPMs; in addition, the wavelength-shifting also better matches the spectral response of commercially available visible-light sensitive SiPMs. However, the total collection efficiency of the liquid argon detector --- without accounting for shadowing from the LEGEND geometry --- is expected to be $\sim 0.1\%$ of the total number of VUV photons produced \cite{gerdacollaborationLiquidArgonLight2022}.

Characterizing the photon detection efficiency (PDE) of SiPMs at cryogenic temperatures is important for modeling the LAr scintillation detector response of LEGEND-200 --- the current phase of LEGEND that is actively taking data. The PDE is the probability of a SiPM detecting a single incident photon and converting it into a useful signal. The overall light collection efficiency of the LAr detector is the product of individual collection, absorption, transmission, emission, and detection probabilities in the chain of steps a VUV photon takes from scintillation to detection in a SiPM; the PDE is the final link in this concatenation. Cryogenic SiPM PDE measurements will also inform the design of the future phase of LEGEND with 1000~kg of active isotope which aims to reach a discovery sensitivity for $0\nu\beta\beta$ decay beyond $10^{28}$ years. Knowledge of the SiPMs' PDE at liquid argon temperature (87~K) from experimental characterizations is important for optimizing LEGEND-1000's LAr light readout, and for accurately projecting the background \cite{legendcollaborationLEGEND1000PreconceptualDesign2021}. 

Although most measurements of the cryogenic PDE of SiPMs tend to focus on SiPMs that are optimized for cryogenic operation \cite{hamvuv4_2024, Nakarmi_Ostrovskiy_Soma_Retiere_Kharusi_Alfaris_Anton_Arnquist_Badhrees_Barbeau_etal._2020, Pershing_Xu_Bernard_Kingston_Mizrachi_Brodsky_Razeto_Kachru_Bernstein_Pantic_etal._2022}, a number of measurements of the cryogenic PDE of non-cryogenically-optimized SiPMs have been made and are available \cite{Acerbi_Paternoster_Merzi_Zorzi_Gola_2023, collazuolStudiesSiliconPhotomultipliers2011}. Several of these measurements have shown that the PDE consistently drops by 20\% to 30\% at cryogenic temperatures compared to room temperature values across a range of wavelengths \cite{Acerbi_Paternoster_Merzi_Zorzi_Gola_2023, hamvuv4_2024}. The physical mechanism responsible for this decrease is not well understood, but is hypothesized to originate from a decrease in either the quantum efficiency, the avalanche triggering probability, or a combination of the two \cite{Acerbi_Paternoster_Merzi_Zorzi_Gola_2023, collazuolStudiesSiliconPhotomultipliers2011}. A drop in quantum efficiency can be explained by an increase in absorption length, and a drop in avalanche triggering probability can be explained by carrier freeze-out at lower temperatures. The characterization of SiPMs at cryogenic temperatures is important for understanding these effects, but is quite challenging as it requires a dedicated setup that holds only the SiPM at cryogenic temperature while ensuring that other optical properties of the setup do not change. The design of our apparatus cools and maintains the SiPM, at ambient pressure, at 77~K by keeping the SiPM in thermal contact with a simple liquid nitrogen reservoir in a dewar. Other experimental setups rely on vacuum systems to guarantee optical stability, while our approach uses liquid nitrogen boil off to similar effect, as systematic studies in section \ref{sec:results} show. The liquid nitrogen bath only allows us to probe one temperature --- unlike other experiments that use closed-cycle cryostats or heat-exchangers --- close to LEGEND's 87~K operating temperature where the results are the most critical to understanding the impact of a changing PDE on the overall light collection efficiency of LEGEND's liquid argon detector.  

In this paper, we describe the design of a test stand capable of measuring the PDE at cryogenic temperatures. We report a characterization of the cryogenic PDEs of two commercially available SiPMs --- a KETEK PM3325-WB-D0 SiPM and a Hamamatsu S13360-3050CS SiPM --- near liquid argon temperature to determine potential effects on the efficiency of the LEGEND liquid argon scintillation readout system. LEGEND currently uses KETEK PM33100T SiPMs; the KETEK PM3325-WB-D0 was one of the latest available SiPM models from KETEK before the acquisition of SiPM assets by BROADCOM\textsuperscript{\textregistered}. The Hamamatsu SiPM, in contrast, has a higher PDE at the peak emission wavelength of the wavelength-shifting fibers currently used in LEGEND and is of R\&D interest for LEGEND-1000. We find that the cryogenic PDEs at $562$~nm for these two devices exhibit a greater than 20\% decrease from their room temperature PDE values across a broad range of overvoltages.

\section{Experimental Method}
A SiPM characterization test stand has been built at the University of Washington to measure the PDE and other properties, such as the dark count rate and breakdown voltage, at both room temperature and cryogenic temperature.  The $\mathrm{PDE}$ of a SiPM can be broken down into three components \citep{acerbiUnderstandingSimulatingSiPMs2019}, 
\begin{equation}
\mathrm{PDE}(V_{OV}, \lambda, T) = QE(\lambda, T) \times P_t(V_{OV}, \lambda, T) \times FF_{\mathrm{eff.}}(V_{OV}, \lambda)\,.
\label{eq:pde}
\end{equation}
$QE(\lambda, T)$ is the quantum efficiency, $P_t(V_{OV}, \lambda, T)$ is the avalanche triggering probability, and $FF_{\mathrm{eff.}}(V_{OV}, \lambda)$ is effective fill factor. Each of these can depend on the wavelength ($\lambda$) of incoming light; the temperature ($T$) of the SiPM; and/or the overvoltage ($V_{OV}$), which is the difference between the operating reverse bias and the breakdown voltage.

\begin{figure}
    \centering
    \includegraphics[width=\textwidth]{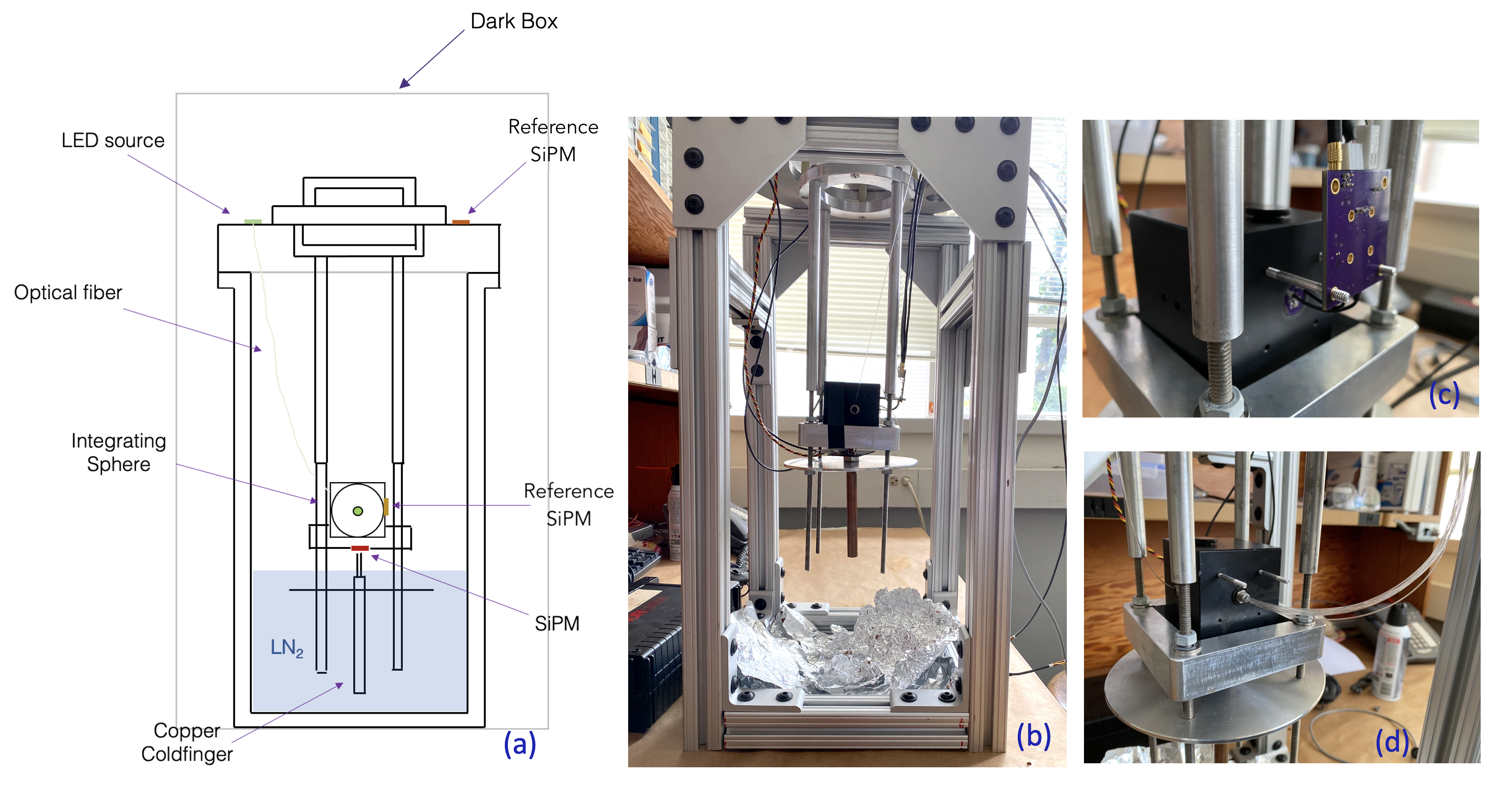}
    \caption{(a) A cross-sectional view of the apparatus used to perform direct and lid cryogenic PDE measurements.  The integrating sphere is suspended above a dewar filled with liquid nitrogen (L$\mathrm{N}_{2}$) and a copper cold finger thermally coupled to the SiPM extends into the L$\mathrm{N}_{2}$ volume. (b) A photo of the integrating sphere suspension set-up that is inserted into the dewar. (c) A photo of the direct PDE reference SiPM configuration with the first stage amplification board at the integrating sphere. (d) A photo of the lid PDE measurement configuration with the optical fiber bundle connected to the integrating sphere.}
    \label{fig:dewar}
\end{figure}

The setup consists of a ThorLabs IS200-4 integrating sphere suspended above a liquid nitrogen bath inside of a dewar contained inside a dark box as shown in Fig.~\ref{fig:dewar}. At each port of the integrating sphere, with perpendicular lines of sight, are the SiPM under test, a BROADCOM AFBR-S4K33C0135L SiPM (the reference SiPM), and a pulsed light emitting diode (LED) light source peaked at 562~nm with a full width at half maximum of 11~nm \cite{Thorlabs}. This wavelength was chosen because it is within LEGEND's WLS fibers' emission spectrum, approximately 460~nm to 580~nm \cite{Wiesinger}, and was readily at hand. One NanOptics\textsuperscript{TM}(030UC-00S) clear, double-cladded optical fiber connects the LED light source ---located outside of the dewar --- to the integrating sphere. The SiPM under test sits on top of a copper cold finger, which allows it to be coupled thermally to a liquid nitrogen bath. The cold finger also ensures that the integrating sphere and the surface of the SiPM under test are not submerged in liquid nitrogen, which would alter the transmissivity and prohibit a clear measurement of the intrinsic PDE; however, future measurements should be done to measure the transmissivity in liquid argon to determine its impact on the overall liquid argon detector efficiency. 

The uncalibrated BROADCOM SiPM is used as a reference for the relative light level in the integrating sphere because an estimate of its room temperature PDE is available from the manufacturer and is constant across measurements. For room temperature PDE measurements, the reference diode is mounted directly to the integrating sphere; for cryogenic PDE measurements, the reference SiPM is placed at the lid of the dewar, thermally isolated from the LN, and optically coupled to the integrating sphere by a bundle of clear fibers. The SiPM signals are read out when triggered by an Agilent 33521 pulse generator that pulses the LED. The SiPM signals are pre-processed first with Texas Instruments LMH6629 transimpedance amplifiers near the integrating sphere and then with pre-amps consisting of two Analog Devices AD8014 amplifiers in series outside the dewar, based on the amplifier found in \cite{DIncecco_Galbiati_Giovanetti_Korga_Li_Mandarano_Razeto_Sablone_Savarese_2018}. The Texas Instruments LMH6629 amplifier near the integrating sphere increases the signal-to-noise ratio so that single-photon resolution can be achieved during room temperature measurements.

The PDE at liquid nitrogen temperature is calculated by taking the ratio of the number of detected photons at the SiPM ($N_{SiPM}$) and the incoming flux of photons at the SiPM port ($\Phi_{SiPM}$) multiplied by the SiPM area ($A_{SiPM}$). Calculation of $N_{SiPM}$ is done using Eqn. \ref{eq:sipm_trigger} and is detailed in section \ref{sec:methods}. We are using the reference SiPM to calculate $\Phi_{SiPM}$, and because the reference SiPM has not been absolutely calibrated (we are taking its value from the manufacturer's datasheet), the PDE we measure is relative to the reference and is given by

\begin{equation}
    \mathrm{PDE}^{\mathrm{rel.}}(V_{OV}, \lambda, 77~\mathrm{K}) = \frac{N_{SiPM}(V_{OV}, \lambda, 77~\mathrm{K})}{\Phi_{SiPM}A_{SiPM}}. \label{eq:ln_pde}
\end{equation}
The incident flux on the test SiPM port is assumed to be equal to the incident flux on the reference SiPM port. During liquid nitrogen PDE measurements, the reference SiPM is placed at the lid of the dewar and is optically coupled to the integrating sphere by a bundle of the clear, double-cladded optical fibers, in order to preserve the known response ($\mathrm{PDE}^{\mathrm{ref.}}$) of the reference SiPM. Because the LED's emission spectrum, $\rho(\lambda)$, spans a range of wavelengths, the reference SiPM's PDE should be a weighted average of the PDE across the LED's emission spectrum, 
\begin{equation}
\mathrm{PDE}^{\mathrm{ref.}} = \int\frac{\lambda}{h c}\left[\frac{\rho(\lambda)}{\eta(\lambda)}\right]\, d\lambda\,.
\label{eq:spectral_response}
\end{equation}
$\eta(\lambda)$ is the spectral response of the reference SiPM; both $\eta(\lambda)$ and $\rho(\lambda)$ are available from the manufacturer. The reference SiPM is operated at a constant applied overvoltage of 5~V. The incident flux on the test SiPM is therefore expressed as

\begin{equation}
    \Phi_{SiPM} = \frac{N_{\mathrm{ref.}}}{A_{\mathrm{ref.}}\mathrm{PDE}^{\mathrm{ref.}}f}, \label{eq:sipm_flux}
\end{equation}
where $N_{\mathrm{ref.}}$ is the number of photons detected by the reference SiPM during the liquid nitrogen temperature measurement, $A_{\mathrm{ref.}}$ is the area of the reference SiPM, and $f$ is the flux ratio that corrects for the transmission coefficient to the reference SiPM (including, but not limited to, optical fiber transmission and coupling to the fibers, which are measured in Section.~\ref{sec:results}) and any increased reflectivity in the sphere due to the fibers. The equality in Eq.~\ref{eq:sipm_flux} assumes that light is distributed equally to both integrating sphere ports, that the transmission coefficient does not vary with temperature, and that the SiPMs are aligned perfectly normal to the port. These assumptions are verified in Section.~\ref{sec:results}. The flux ratio can be measured by first comparing the number of photons measured by the reference SiPM at room temperature in this lid PDE configuration to the number of photons measured at room temperature when the reference SiPM is placed directly on the sphere. This is then normalized by the number of photons detected by the SiPM under test in order to correct for any fluctuations in the LED's brightness or increased reflectivity from the fibers, 

\begin{equation}
    f = \frac{N_{\mathrm{ref.}}^{\mathrm{lid}}}{N_{\mathrm{ref.}}^{\mathrm{dir.}}}\frac{N_{\mathrm{SiPM}}^{\mathrm{dir.}}}{N_{\mathrm{SiPM}}^{\mathrm{lid}}}. \label{eq:flux_ratio}
\end{equation}
Here $N_{\mathrm{ref.}}^{\mathrm{dir.}}$ and $N_{\mathrm{SiPM}}^{\mathrm{dir.}}$ are measured when both SiPMs are mounted directly on the integrating sphere, while $N_{\mathrm{ref.}}^{\mathrm{lid}}$ and $N_{\mathrm{SiPM}}^{\mathrm{lid}}$ are measured with the test SiPM on the integrating sphere and the reference SiPM at the lid. These four quantities are measured at room temperature and the value of $f$ is of order $\sim 1$.

\section{Data Acquisition and Processing \label{sec:methods}}
The signal output of the SiPMs were recorded by a CAEN DT5730 digitizer with a 500 MS/s sampling rate, 14-bit resolution, and dynamic range of 2 Vpp.  The data were acquired using CAEN CoMPASS software \cite{CAEN_2023}. The LED was pulsed with a 16 ns pulse width and a 1 kHz frequency from the function generator. Waveforms of length 10 $\mu$s, with the trigger at $1.5 \mu$s, were acquired. These waveforms were long enough to ensure that the last microsecond of data --- which is several microseconds after the LED has turned off and the SiPM has recharged --- is taken in dark conditions so that a dark count rate could be measured. Three sets of waveform data were taken with 90~s acquisition time at each overvoltage in order to improve statistics. This process of collecting waveform data at each overvoltage was done with the direct PDE configuration at room temperature, and with the lid PDE configuration at room and liquid nitrogen temperatures. Further dark data were taken to enable a full characterization of the two SiPMs, see appendix~\ref{sec:dark_characterization}.

To ensure that the PDE measurements were not biased by correlated noise such as cross-talk, the pedestal method of measuring the PDE was used as described in \cite{Otte_Hose_Mirzoyan_Romaszkiewicz_Teshima_Thea_2006}. To summarize, waveforms were acquired synchronously by triggering on the pulser so that baseline events, where the SiPM did not detect photons, were recorded. If the number of detected photons is Poisson distributed, then the probability of not detecting any photoelectrons, $P(N_{\mathrm{detected}}=0)$, gives an unbiased estimate of the Poissonian mean ($\overline{N_\gamma}$) because the baselines are unbiased by any correlated noise \cite{Otte_Hose_Mirzoyan_Romaszkiewicz_Teshima_Thea_2006}:
\begin{equation}
    P(N_{\mathrm{detected}}=0) = \exp\left(-\overline{N_\gamma}\right).
\end{equation}
The probability of not detecting any photons is equal to the number of baseline events ($N_0$) divided by the total number of LED pulses ($N_{\mathrm{total}}$). In this way, the mean number of detected photons becomes 
\begin{equation}
    \overline{N_\gamma} = -\ln\left(\frac{N_0}{N_{\mathrm{total}}}\right).
\end{equation}

\begin{figure}
    \centering
    \includegraphics[]{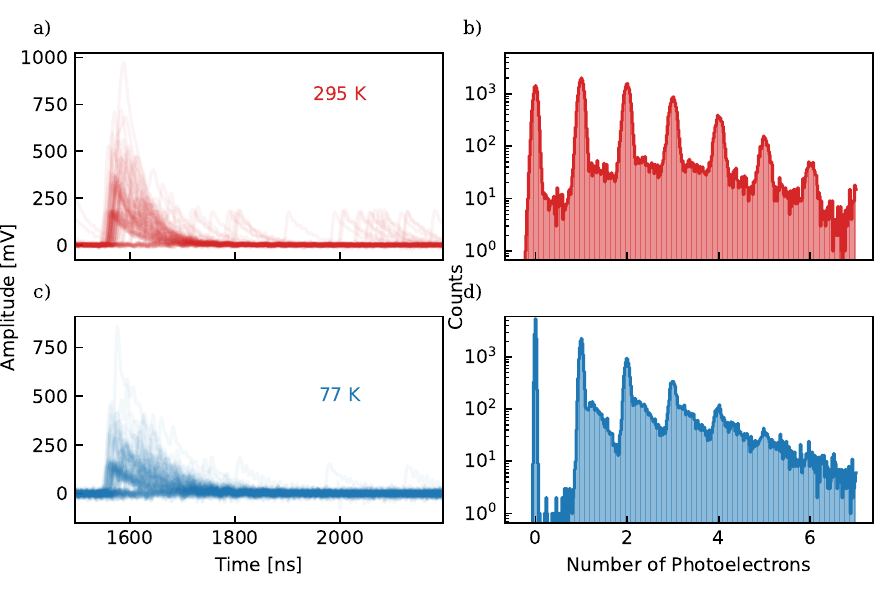}
    \caption{Hamamatsu SiPM waveforms at 2.5~V overvoltage, triggered synchronously with the LED pulse, for room temperature (a) and liquid nitrogen temperature (c). Normalized charge spectra are used to calculate the mean number of photons detected and are shown at room temperature (b) and liquid nitrogen temperature (d), where the peak shape for $N_{>0}$ of Hamamatsu 77~K data was found to be distorted by after pulsing, see Appendix.}
    \label{fig:sipm_waves}
\end{figure}

To get the number of baseline events $N_0$ the SiPM signals are first integrated in a 1 $\mu$s window around the LED trigger time as well as the last 1 $\mu$s of the waveform to get the dark count rate. Examples of the synchronously triggered waveforms are shown in Fig.~\ref{fig:sipm_waves}a. When the histograms of these integrated signals are made, the peaks that correspond to the integral of the baseline events are called the pedestal peaks, as can be seen in Fig.~\ref{fig:sipm_waves}b. The light level of the pulsed LED was set low enough so that an appreciable number of baseline events were acquired by making sure the pedestal peak was resolvable by eye when using the CoMPASS online energy spectrum. This corresponds to roughly $5.73\pm 0.08$ photons per pulse incident on the SiPM face, as measured by the reference diode in the direct PDE configuration of the test stand. A Gaussian is then fit to the pedestal peaks in the histograms to determine the pedestal peak locations $\mu$ and standard deviations $\sigma$. The number of baseline events $N_0$ in light and dark conditions are then determined by the number of counts below a $\mu+3\sigma$ threshold. This routine is necessary because differences in the electronics noise between room and liquid nitrogen temperatures preclude the use of a single threshold. To get the mean number of photons detected by either the test SiPM or the reference SiPM ($N_{\gamma}$) requires subtraction of the dark count rate,
\begin{equation}
N_{\gamma} = \left[ -\ln\left(\frac{N_{0, \mathrm{light}}}{N_{\mathrm{total}}}\right) +\ln\left(\frac{N_{0, \mathrm{dark}}}{N_{\mathrm{total}}}\right)\right]
\,.
\label{eq:sipm_trigger}
\end{equation} 
$N_{0, \mathrm{dark}}$ is the number of baseline events in dark conditions, $N_{0, \mathrm{light}}$ is the number of baseline events synchronous with the LED pulse. The total number of pulses is the same for both light and dark because the dark counts come from the portion of the waveform long after the LED turns off. 

Post-processing of the data from CAEN CoMPASS is done with custom open source software, \texttt{SiPM Studio}\footnote{https://github.com/SamuelBorden/sipm\_studio}. This software parallel processes PDE data and automatically performs the Gaussian fits to the pedestal peak.

\section{Results} \label{sec:results}

\begin{figure}
    \centering
    \includegraphics[]{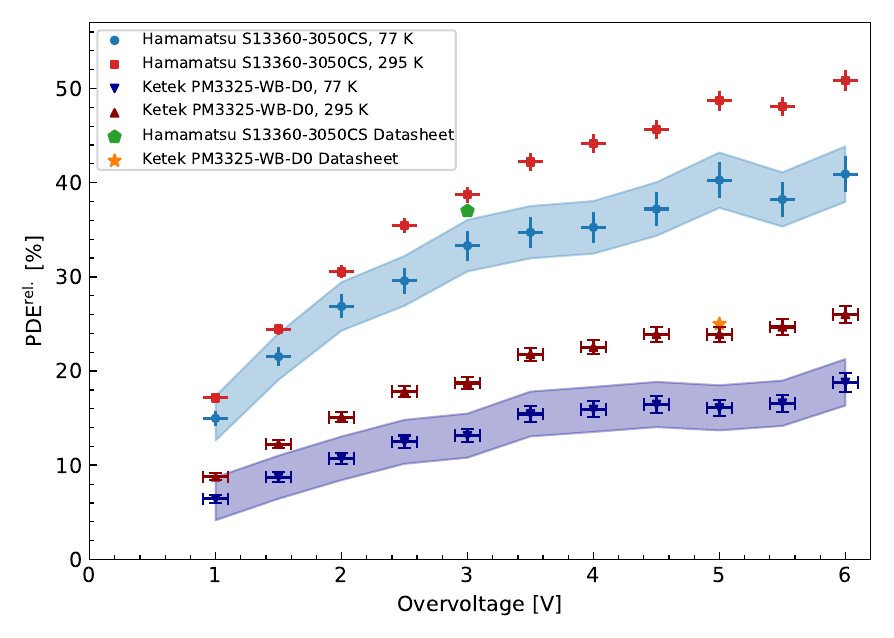}
    \caption{Measurement of the relative PDE for a KETEK PM3325-WB-D0 SiPM and a Hamamatsu S13360-3050CS SiPM for a range of overvoltages. Measurements are done at room temperature and liquid nitrogen temperature using $562$~nm light. The datasheet values for the PDEs at the peak LED wavelength used are provided for reference. The error bands correspond to combined statistical and systematic (2\% from the flux ratio $f$ and 1\% from $\mathrm{PDE}^{\mathrm{ref.}}$) uncertainties.}
    \label{fig:pde_measure}
\end{figure}

The relative PDE was measured at room temperature and at liquid nitrogen temperature for the KETEK SiPM and Hamamatsu SiPM, see Fig.~\ref{fig:pde_measure}. Errors on all three quantities in Eq.~\ref{eq:ln_pde} were propagated during error calculations. In order to account for a $\pm 0.1$~V uncertainty on the breakdown voltage of the reference SiPM, a systematic uncertainty on $\mathrm{PDE}^{\mathrm{ref.}}$ of $1\%$ was found from the manufacturer's datasheet and was used in the analysis. The room temperature was 72~$^\circ$F (295~K) held by an air conditioning unit and the liquid nitrogen temperature was measured to be 77~K with a calibrated PT1000 sensor placed in between the SiPM board and copper cold finger. These temperatures were stable within 1~K during data taking. The breakdown voltage of the KETEK SiPM was measured at room temperature to be 24.1~V; the breakdown voltage at 77~K was 20.6~V. The breakdown voltages were obtained by extrapolating the SiPM gain, $G$, as a function of reverse bias to $G=1$ and were confirmed with a reverse IV curve from a Keithley 2450 Sourcemeter\textsuperscript{\textregistered}. Likewise, the breakdown voltage of the Hamamatsu SiPM was measured at room temperature to be 51.5~V; the breakdown voltage at 77~K was 42.2~V. Because the breakdown voltage depends on the temperature, the breakdown voltage was also used as a cross-check on the temperature. The breakdown voltages at liquid nitrogen temperature were determined by submerging the devices completely in liquid nitrogen in a separate experiment prior to the PDE measurement. While the reference SiPM was mounted on the lid, the breakdown voltage of the reference SiPM was found to not change after an LN fill, confirming that it was kept at room temperature.

The PDE was measured for a range of overvoltages and results show a significant decrease in PDE at liquid nitrogen temperature for all overvoltages. This decrease in PDE is present even when accounting for possible changes in the optical properties of the setup at different temperatures or between configurations. Any systematic impact of the light uniformity in the integrating sphere or SiPM alignment would be seen during room temperature PDE measurements. Our room temperature PDE measurements are in good agreement with the manufacturer's datasheets, as seen in Fig.~\ref{fig:pde_measure}. Moreover, the agreement between manufacturer and measured room temperature PDE values indicates that all three SiPMs are consistent with their manufacturer reported values, within our experimental uncertainties; although, a common systematic offset between all three cannot be excluded. Any differences in SiPM alignment when the reference diode is moved to the lid for lid configuration PDE measurements are accounted for in measurements of the flux ratio, Eq. \ref{eq:flux_ratio}. Systematic studies were performed to determine if there was any temperature dependence in the transmission coefficient of the optical fibers connecting to the reference SiPM, the emission spectrum of the LED, or the reflectivity of the integrating sphere.

\begin{figure}
   \centering
    \includegraphics[]{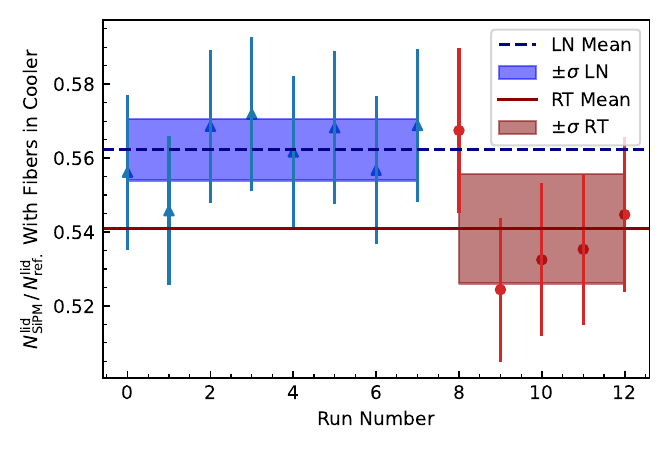}
    \caption{The fluctuation in the effective PDE, $\frac{N_{\mathrm{SiPM}}^{\mathrm{lid}}}{N_{\mathrm{ref.}}^{\mathrm{lid}}}$, when only the optical fibers were cooled down sets a limit on the systematic error at $2\%$.}
    \label{fig:fiber_transmission}
\end{figure}

We characterized the temperature dependence of the optical fibers' transmission coefficient (and the flux ratio $f$) by taking the dewar insert as shown in Fig.~\ref{fig:dewar}(d) and running $25$~cm of the $30$~cm long fibers through a Styrofoam cooler that was then filled with liquid nitrogen. The fibers were thus the only part of the test stand that was cold. One end of the fibers was connected to the integrating sphere, with the Hamamatsu SiPM placed at the opposite port, while the other end was coupled to the reference SiPM. The integrating sphere, the Hamamatsu SiPM, and the reference SiPM were all at room temperature. We then took PDE data with and without liquid nitrogen on the fibers while triggering on LED pulses sent into the integrating sphere. Fig.~\ref{fig:fiber_transmission} shows the effective PDE --- the number of photons detected by the Hamamatsu on the integrating sphere divided by the number of photons detected by the reference SiPM at the end of the fiber --- when the fibers were at room and liquid nitrogen temperatures. The difference between the mean effective PDEs measured over several trials was found to be $2\%$, which places a limit on the systematic error in our cryogenic PDE measurements.

We also investigated the systematics arising from the finite width of the LED's emission spectrum. Using Eq.~\ref{eq:spectral_response} to calculate the reference SiPM PDE instead of using the reference SiPM's PDE at the peak emission wavelength of the LED introduces only an average $0.02\%$ difference in our final cryogenic PDE results, which is below our statistical sensitivity. We therefore compute the cryogenic PDE using only the peak emission wavelength of the LED.

\begin{figure}
   \centering
    \includegraphics[]{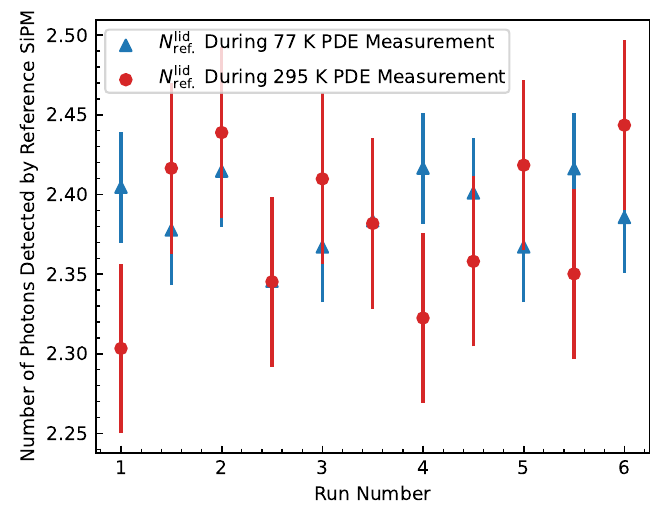}
    \caption{Comparing the number of photons detected by the reference SiPM located at the lid during room and cryogenic temperature PDE measurements shows that there are no systematic shifts in the LED's emission spectrum or the transmission coefficient of the optical fibers.}
    \label{fig:reference_photons}
\end{figure}

We checked that $\rho(\lambda)$ (the LED's emission spectrum) and the integrating sphere's optical properties were not functions of temperature. While the LED emission spectrum should not change because the LED is placed outside of the dewar, the optical fiber that connects the LED to the integrating sphere could have a changing transmission coefficient as a function of temperature. To constrain this possibility, we compared the number of photons detected at the reference SiPM at 77~K and 295~K while the test stand was in the lid PDE measurement configuration. The reference SiPM has a comparable spectral response to the SiPMs under test. The results, shown in Fig.~\ref{fig:reference_photons}, indicate that the reference SiPM detects roughly the same number of photons when the test stand is at both room and cryogenic temperatures. Thus there is neither a shift in the emission spectrum of the LED nor a shift in the reflectivity of the integrating sphere. This measurement also serves as an \textit{in situ} cross-check that the optical fibers' transmission probability is not temperature dependent. 

We ensured that condensation was not forming inside the integrating sphere or on the surfaces of the SiPMs. First, prior to data taking, a liquid nitrogen fill was performed and the dewar insert was lifted to visually check that no condensation was present. Second, the number of photons detected by the reference SiPM at the lid does not change significantly between room temperature and cryogenic temperature measurements in Fig.~\ref{fig:reference_photons}: any change in reflectivity engendered by condensation inside the integrating sphere would cause a systematic change in the number of photons detected by the reference SiPM. The nitrogen gas boil-off from the liquid nitrogen ensures that condensation cannot form. Therefore, any changes in the optical properties of the test stand between the two temperatures do not account for the observed decrease in PDE.

The decrease in PDE at 77 K indicates that some combination of the quantum efficiency, avalanche triggering probability, and effective fill factor may be changing with temperature. Similar reductions of the PDE at cryogenic temperatures have been found in other SiPMs at 10 K using a calibrated system \cite{Ma_Zhou_Gu_Liao_Gao_Peng_Zheng_An_Zhang_Zhang_etal._2023}, and at 77 K using uncalibrated systems \cite{Acerbi_Paternoster_Merzi_Zorzi_Gola_2023, collazuolStudiesSiliconPhotomultipliers2011, Nakarmi_Ostrovskiy_Soma_Retiere_Kharusi_Alfaris_Anton_Arnquist_Badhrees_Barbeau_etal._2020, Biroth_Achenbach_Lauth_Thomas_2016} and calibrated systems \cite{hamvuv4_2024, Jamil_Ziegler_Hufschmidt_Li_Lupin-Jimenez_Michel_Ostrovskiy_Retiere_Schneider_Wagenpfeil_etal._2018}. These systems show similar decreases in PDE of roughly 30\% at liquid nitrogen temperatures for wavelengths close to 560~nm. An FBK NUV-HD-cryo SiPM operated at 3~V overvoltage was measured to have a PDE drop from $\sim$24\% at room temperature to $\sim$18\% at 75~K for 525~nm light; however, the decrease in PDE at 5~V overvoltage is much smaller, dipping from $\sim$34\% down to only $\sim$29\%  \cite{Acerbi_Paternoster_Merzi_Zorzi_Gola_2023}. A VUV-sensitive Hamamtsu SiPM at 4~V overvoltage displayed a similar decrease in PDE from 28.4\% at room temperature to 20.1\% at 77~K for 570 nm light as reported in \cite{hamvuv4_2024}. Because silicon is an indirect band-gap semiconductor and fewer phonons are present in the crystal lattice at liquid nitrogen temperature, the probability of charge carriers moving to the conduction band from the absorption of a photon will be lower. This is reflected in the increase of the absorption depth of photons at a given energy at liquid nitrogen temperatures; this effect could reduce the quantum efficiency \cite{Acerbi_Paternoster_Merzi_Zorzi_Gola_2023, collazuolStudiesSiliconPhotomultipliers2011}. The band-gap of silicon also increases with decreasing temperature, but this may produce negligible effects on the quantum efficiency \cite{bludauTemperatureDependenceBand1974}. In addition, the recombination probability for electron-hole pairs before reaching the avalanche region may be higher at liquid nitrogen temperature. Similar findings were observed when researching the quantum efficiency of silicon thin film solar cells \cite{wagnerTemperaturedependentQuantumEfficiency2003}. The avalanche triggering probability could also decrease due to carrier loss from carrier freeze-out \cite{collazuolStudiesSiliconPhotomultipliers2011}; however, the triggering probability may also increase slightly due to larger carrier mobility at liquid nitrogen temperatures. More work needs to be done to disentangle the many effects that might be causing the decrease in the PDE. It is not expected that thermal expansion significantly impacts the effective fill factor because the thermal expansion coefficient for silicon at liquid nitrogen temperatures amounts to only a 0.02\% change in area, which is too small to explain the observed decrease in PDE. 

\section{Conclusions}
An experimental setup has been built at the University of Washington to characterize SiPMs at liquid nitrogen temperature to assess the efficiency of the liquid argon veto readout system to be deployed in LEGEND-1000. The SiPM test stand consists of an integrating sphere suspended above a liquid nitrogen bath in a dewar housed inside a dark box. The SiPM is in thermal contact with a copper cold finger that is partially submerged in liquid nitrogen and the integrating sphere is in a room-temperature thermal bath. Numerous systematics studies were performed to ensure that no other optical properties of the experimental apparatus changed at liquid nitrogen temperatures. Measurement of both a KETEK PM3325-WB-D0 SiPM and Hamamatsu S13360-3050CS SiPM PDE at liquid nitrogen temperature suggests the liquid argon veto readout system has a lower efficiency compared to the expected PDE for room temperature operation. We expect the same behavior to persist at liquid argon temperatures because they are only $10$ K away from the measured liquid nitrogen temperatures: past studies have shown that the PDE at $87$~K for similar wavelengths is only $1.1$ times larger than the PDE at $77$~K \cite{collazuolStudiesSiliconPhotomultipliers2011}. The reduction in PDE seems to be caused by temperature dependencies to the intrinsic quantum efficiency and avalanche triggering probability of the SiPM.  Changes to the effective fill factor may be present but are most likely negligible in comparison to the changes in quantum efficiency and avalanche triggering probability. Additional measurements of the PDE at different wavelengths at cryogenic temperatures should be pursued. The measurement of a decrease in PDE greater than 20\% for all overvoltages at cryogenic temperatures for a SiPM sensitive to the visible spectrum is significant information for assessing and improving the background suppression capability of LEGEND-1000. 

\acknowledgments{
This work was funded by the Department of Energy Office of Nuclear Science. The authors appreciate the technical assistance and advice of J.F.~Amsbaugh, T.H.~Burritt, G.~Giovanetti, G.~Holman, A.~Hostiuc, C.J.~Nave, D.~Peterson, G.~Song, T.D.~Van Wechel, L.~Varriano, I.~Wang, and C. Wiseman. The authors would also like to thank A.~de~St.~Croix for helpful discussion.}

\bibliographystyle{JHEP}
\bibliography{main}

\appendix 
\section{Dark Characterization}\label{sec:dark_characterization}
To estimate the SiPM dark count behavior, we took additional data under no LED illumination while the digitizer triggered on waveforms close to the 0.5 photoelectron amplitude. Waveforms with 10~$\mu$s long traces were acquired for 60 seconds at each reverse bias at room temperature; however, at liquid nitrogen temperatures, an hour of data per reverse bias was taken to ensure high enough statistics. The custom software processed dark data by Wiener filtering the waveforms to get a more precise determination of dark pulses' amplitude and arrival time. Wiener filtering allows for an unbiased estimate of pulse height even when a pulse occurs on the decay tail of a previous pulse. 

\begin{figure}
    \centering
    \includegraphics[width=\textwidth]{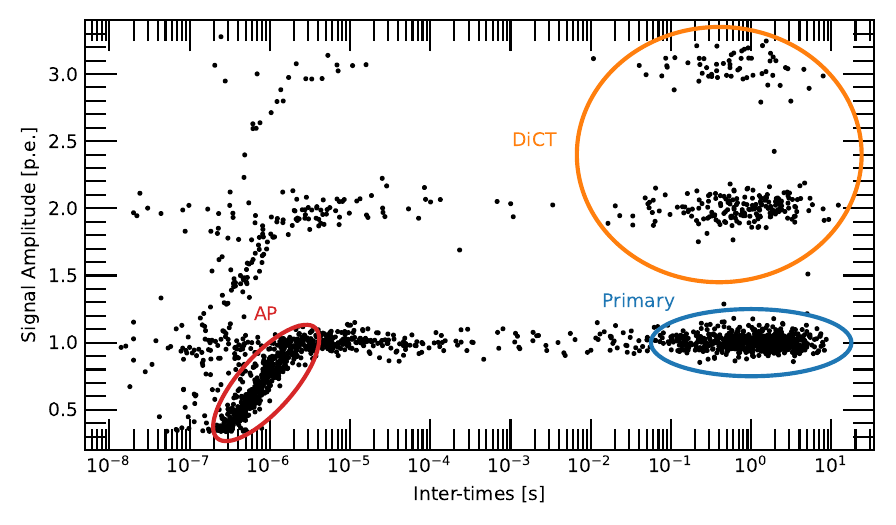}
    \caption{The different dark count populations shown as waveform amplitude vs waveform inter-time spacing for a 4 V overvoltage KETEK at liquid nitrogen temperature. Inter-time is defined as the time to the next waveform the triggered waveform.  Primary dark counts are produced from thermal excitations, DiCT is produced from a SPAD triggering one or more neighboring SPADs simultaneously, and AP is from the trapping and re-releasing of charge carriers that generate an avalanched while the SPAD is not fully quenched.}
    \label{fig:dark_counts}
\end{figure}

Under no illumination, SPADs can trigger an avalanche from thermal excitations due to the environment's ambient temperature along with other correlated noise: direct cross-talk (DiCT), delayed cross-talk (DeCT), and afterpulsing (AP). The number of thermal excitations observed within a timing window is described by a Poisson distribution, thus the probability density function of inter-time spacings is an exponential distribution
\begin{equation}
P_{\tau}(t) = \frac{1}{\tau}\exp\left(\frac{-t}{\tau}\right)\, ,
\end{equation}
where $1/\tau$ is the primary dark count rate.   Secondary photons generated in an avalanche have some probability of triggering another avalanche in a neighboring SPAD simultaneously, which causes a waveform that is an integer multiple times larger than the single photoelectron signal, called direct cross-talk (DiCT).  The DiCT probability is calculated as the ratio of events beyond the 1.5 photoelectron threshold to that beyond the 0.5 photoelectron threshold.  It is also possible for the cross-talk to be delayed if the secondary photons create charge carriers that diffuse towards the avalanche region on the 10 ns time scale, called delayed cross-talk (DeCT). DeCT was not observed to be a significant contribution to dark counts from the KETEK SiPM or Hamamatsu SiPM.  Afterpulsing (AP) is the probability a trapped charge carrier during an avalanche re-releases and triggers an additional avalanche when the SPAD is not fully quenched.  The afterpulsing probability is calculated by dividing the total number of afterpulsing events by the total number of waveforms. The dark count populations for liquid nitrogen temperature are shown in Fig.~\ref{fig:dark_counts}.

\begin{figure}
    \centering
    \includegraphics[width=\textwidth]{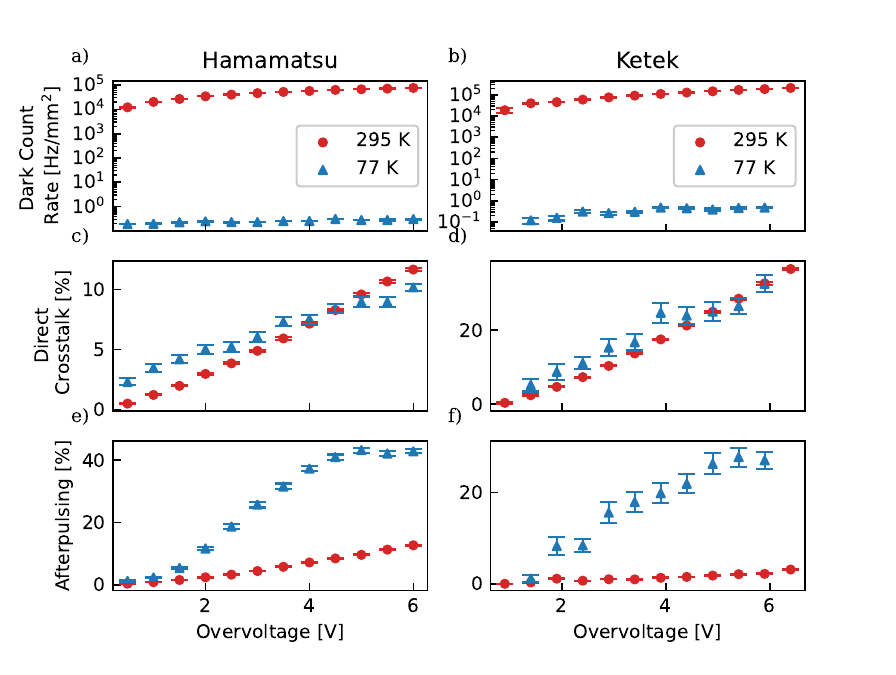}
    \caption{Dark characterization data for the KETEK PM3325-WB-D0 and Hamamatsu S13360-3050CS SiPMs at room and liquid nitrogen temperatures as a function of overvoltage. Room temperature data for the Ketek were taken with a $0.1$~V offset from the breakdown voltage, so liquid nitrogen data were acquired with the same offset. In addition, the number of events for the Ketek SiPM at $0.4$~V overvoltage at liquid nitrogen temperature was too low to analyze due to the extremely low dark count rate, so the data point was omitted.}
    \label{fig:dark_plot}
\end{figure}

The dark count rate was measured for several overvoltages in both devices at both room temperature and liquid nitrogen temperature. The dark count rate is suppressed by several orders of magnitude at cryogenic temperatures, as seen in Fig.~\ref{fig:dark_plot}(a, b). This is to be expected as the thermal generation of carriers follows a modified Arrhenius relation --- thermal generation is suppressed so much that, at liquid nitrogen temperatures, band-to-band tunneling dominates \cite{collazuolStudiesSiliconPhotomultipliers2011}.

The direct cross-talk was also measured at a variety of overvoltages for both devices. The cross-talk probability is larger at cryogenic temperatures for both devices, but there is a cross-over point where it becomes equal to the room temperature cross-talk probability, as seen in Fig.~\ref{fig:dark_plot}(c) and Fig.~\ref{fig:dark_plot}(d). This is somewhat surprising as the cross-talk probability depends on the gain (via the number of secondary photons produced) and the avalanche triggering probability. The gain does not change significantly at cryogenic temperatures, but the avalanche triggering probability might, as was discussed in Section~\ref{sec:results} \cite{collazuolStudiesSiliconPhotomultipliers2011}. This could be evidence for an increased avalanche triggering probability arising from the larger mobility of charge carriers at liquid nitrogen temperatures \cite{Acerbi_Paternoster_Merzi_Zorzi_Gola_2023}. The cross-over could indicate that the avalanche triggering probability saturates faster as a function of overvoltage in the liquid nitrogen case. The slight temperature dependence of the cross-talk probability agrees with the elevated cross-talk probability at liquid nitrogen temperatures reported in \cite{Acerbi_Davini_Ferri_Galbiati_Giovanetti_Gola_Korga_Mandarano_Marcante_Paternoster_etal._2017}. The difference between room and liquid nitrogen temperature cross-talk probabilities is also not as substantial as it is for the afterpulsing.

The Hamamatsu and KETEK afterpulsing rates are significantly higher at liquid nitrogen temperatures, as shown in  Fig.~\ref{fig:dark_plot}(e,f). A higher afterpulsing rate is potentially explained by new charge traps or by the increased charge trapping lifetime at cryogenic temperatures: trapped carriers are more likely to be re-released while the cell is recovering and are thus more likely to trigger an afterpulse \cite{Gola_Acerbi_Capasso_Marcante_Mazzi_Paternoster_Piemonte_Regazzoni_Zorzi_2019}. Carrier freeze-out could lead to the activation of new shallow charge traps that could explain both the high afterpulsing rate as well as the decrease in PDE via a decrease in the avalanche triggering probability due to carrier loss \cite{collazuolStudiesSiliconPhotomultipliers2011}. In addition to charge trapping re-releases, afterpulsing can also occur due to secondary carriers drifting from the substrate into the recharging SPAD \cite{Acerbi_Ferri_Zappala_Paternoster_Picciotto_Gola_Zorzi_Piemonte_2015}. This type of afterpulsing could also increase due to increased mobility of charge carriers at cryogenic temperatures and longer SPAD recovery times: carriers in the bulk silicon are more easily drifted into the active region and trigger an afterpulse \cite{Tobehn-Steinhauser_Reiche_Schmelz_Stolz_Frohlich_Ortlepp_2021}. Fig. \ref{fig:sipm_waves}d shows that the higher afterpulsing rate degrades the Hamamatsu charge spectrum at liquid nitrogen temperatures. A similarly high cryogenic afterpulsing rate for a Hamamatsu device was recently reported in \cite{Guarise_Andreotti_Calabrese_Cotta_Ramusino_Cicero_Fiorini_Giammaria_Lax_Luppi_Minotti_etal._2021}.

\end{document}